\documentstyle[prb,aps,epsfig,multicol]{revtex}
\begin{document}
\sloppy
\draft
\title
{Bose-Einstein condensation in tight-binding bands}
\author{R. Ramakumar$^{a,}$\cite{add2} and A. N. Das$^{a}$}
\address{$^{a}$Condensed Matter Physics Group, 
Saha Institute of Nuclear Physics,
1/AF Bidhannagar, Kolkata-700064, INDIA \\
$^{b}$Department of Physics and Astrophysics, University of Delhi,
Delhi-110007, INDIA} 
\date{9 July 2005}
\maketitle
\begin{abstract}
  We present a theoretical study of 
condensation of bosons in tight binding bands corresponding
to simple cubic, body centered cubic, and
face centered cubic lattices. We have analyzed
non-interacting bosons, weakly interacting bosons using Bogoliubov
method, and strongly interacting bosons through a renormalized
Hamiltonian approach valid for 
number of bosons per site less than or equal to unity. In 
all the cases studied, we find that
bosons in a body centered cubic lattice has the highest Bose condensation
temperature. The growth of condensate fraction of non-interacting
bosons is found to be very close to that of free bosons. The
interaction partially depletes the condensate 
at zero temperature and close to it,
while enhancing it beyond this range below the
Bose-Einstein condensation temperature.
Strong interaction enhances
the boson effective mass as the band-filling is increased and
eventually localizes them to form a Bose-Mott-Hubbard insulator
for integer filling. 
\end{abstract}
\pacs{PACS numbers: 03.75.Lm, 03.75.Nt, 03.75.Hh, 67.40.-w}
\begin{multicols}{2}
\narrowtext
\section{Introduction}
\label{sec1}
In a many boson system, when the thermal de-Broglie wavelength
of a particle becomes comparable to the inter-particle separation, a
condensation in momentum space occurs at a finite temperature
and a macroscopic number of particles occupy  the lowest
single particle energy level and enter in to a phase locked state. 
This phenomenon predicted by
Einstein\cite{einstein} by applying Bose statistics\cite{bose}
to a three dimensional homogeneous system of non-interacting atoms
in the thermodynamic limit is well known as 
the Bose-Einstein condensation.
Though it took seventy years to eventually observe 
Bose-Einstein condensation\cite{cornell,davis,bradley,klep}
in metastable, inhomogeneous, finite, and  three 
dimensional Bose atom vapors, progress
ever since has been tremendous. Extensive investigations
of various aspects of the condensate,
a macroscopic quantum coherent state of atoms,
is now a rapidly expanding field 
of research\cite{pita,leg,ketterle,pethick,hall}.
\par
One of the interesting recent developments in this field
is the observation\cite{greiner}, by Greiner and collaborators, of a 
transition between superfluid and
Mott insulating phases of bosons in an optical lattice.
Experimentalists have explored
this superfluid-Mott transition
and the excitation spectra in the superfluid\cite{greiner,schori} and Mott insulating phases\cite{greiner}.
In these experiments, the condensate is adiabatically transferred to a
simple cubic optical lattice produced by counter propagating laser beams.
By changing the
characteristics of the laser beams, it has been possible to
achieve great control over 
$t/U$, where $t$ is the inter-site hopping energy and
$U$ the
on-site boson-boson interaction energy. It has been shown that
bosons in optical lattices can be adequately modeled employing
a clean Bose-Hubbard model\cite{jak}. That a band-width controlled transition from 
a superfluid state to a Bose-Mott-Hubbard (BMH) insulating state
is possible at a critical value of $t/U$ for integer number
of bosons per site was predicted
in theoretical 
studies\cite{jak,fisher,nandini,shesh,monien,oost}
on Bose-Hubbard model.
In the BMH insulator, the bosons are site localized and the 
single particle excitation
spectrum acquires a gap. 
Recently, superfluid to Mott insulator transition was observed\cite{stof} in
finite one-dimensional optical lattices as well.  
\par
The experimental realization of bosons in optical lattices
provides a microscopic laboratory for the exploration of
collective behavior of quantum many particle systems
in narrow energy bands with great control on $t$, $U$, and the number
of boson per site ($n$). There is
already theoretical studies\cite{black,baizakov} on the possibility of creating
different types of two dimensional lattices (triangular, square,
hexagonal, for example). It has been proposed\cite{santos}
recently that trimerized
optical Kagome lattice can be achieved experimentally and
that a superfluid-Mott transition at fractional filling is 
possible for bosons in this lattice. It is reasonable
to expect that three and two 
dimensional optical lattices of
different symmetries will be created in near future. 
Many experimental groups have produced three dimensional optical
lattices and experimental studies of Bose condensates 
in these lattices will surely receive increasing attention in
the coming years.
Motivated
by such possibilities, we present a theoretical study of
Bose-Einstein condensation in tight binding
bands corresponding to simple cubic ({\em sc}), body centered
cubic ({\em bcc}), and face centered cubic ({\em fcc}) lattices.
We have analyzed non-interacting, weakly interacting, and strongly
interacting bosons. The weakly interacting bosons were analyzed
using a Bogoliubov type theory\cite{bogo}. For the strongly interacting
bosons, we use a renormalized Hamiltonian valid for $n\leq 1$
obtained by projecting out on-site multiple occupancies. 
This analysis is presented in the next section and the conclusions
are given in Sec. III.
\section{Bose condensation in tight binding bands}
\label{sect2}
\subsection{Non-interacting bosons}
\label{subsec}
In this section we discuss the simplest of the three cases studied.
The Hamiltonian of the non-interacting bosons in a tight-binding 
energy band is:
\begin{equation}
H=\sum_{{\bf k}}[\epsilon({\bf k})-\mu]c^{\dag}_{{\bf k}}c_{{\bf k}}\,,
\end{equation}
where $\epsilon({\bf k})$ is the band-structure corresponding to
{\em sc}, {\em bcc}, and {\em fcc} lattices, $\mu$ the
chemical potential, and $c^{\dag}_{{\bf k}}$ is the boson
creation operator. Confining to nearest
neighbor Wannier functions overlaps, these band structures 
when lattice constant is set to unity are,
\begin{eqnarray}
\epsilon_{sc}(k_{x},k_{y},k_{z})&=&-2t\sum_{\mu=x}^{z}cos(k_{\mu})\,,
\nonumber \\
\epsilon_{bcc}(k_{x},k_{y},k_{z})&=&-8t\prod_{\mu=x}^{z}
cos\left(\frac{k_{\mu}}{2}\right)\,,
\end{eqnarray}
and,
\begin{equation}
\displaystyle
\epsilon_{fcc}(k_{x},k_{y},k_{z})=-2t\sum_{\mu=x;\mu\neq\nu}^{z}\sum_{\nu=x}^{z}
cos\left(\frac{k_{\mu}}{2}\right)cos\left(\frac{k_{\nu}}{2}\right)\,,
\end{equation}
where $t$ is the nearest neighbor boson hopping energy.
The condensation temperature ($T_B$) for bosons in these bands 
can be calculated from the boson number equation:
\begin{equation}
n=\frac{1}{N_{x}N_{y}N_{z}}\sum_{k_{x}}\sum_{k_{y}}\sum_{k_{z}}
\frac{1}{e^{\left[\epsilon(k_{x},k_{y},k_{z})-\mu\right]/k_{B}T}-1}\,,
\end{equation}
where $N_{s}=N_{x}N_{y}N_{z}$ is
the total number of lattice sites,
$k_{B}$ the Blotzmann constant, and $T$ the temperature.
At high temperature, the chemical potential is large and
negative. As the temperature comes down, the chemical
potential raises gradually to eventually hit the
bottom of the band. Below this temperature there is macroscopic
occupation of the band bottom and we have a Bose condensate. 
On further reduction of temperature,
the chemical potential is pinned to the bottom of the band
and bosons are progressively transfered from excited states
in to the condensate. All the particles are in the condensate
at absolute zero temperature. 
Fixing the chemical potential at the bottom of the band the solution of
the number equation gives the Bose condensation temperature ($T_B$). For
$T>T_B$ a lower value of $\mu$ satisfies the number equation, while for
$T<T_B$ the RHS of Eq. (4) is less than 
the number of bosons ($n$), the
difference being the number of condensate particles ($n_0$). We determined
$T_B$ and $n_0$ for different lattices as a function of filling and
temperature. Results of these calculations are shown in Figs. 1-5.
We find that for bosons in the tight binding
band corresponding to the {\em bcc} lattice has the highest
Bose condensation temperature. Comparing the single boson
Density Of States (DOS), we find that the band structure
with smallest DOS near the bottom of the band has the 
highest $T_B$. 
 The physical reason
behind it is that, as the temperature is lowered from above the
Bose condensation temperature, the bosons are transferred from the high
energy states to the low energy states following the Bose distribution
function. 
To accommodate these bosons the chemical potential would touch the 
bottom of the boson band at a higher temperature for a system with smaller
DOS near the bottom of the band compared to a system which has larger
DOS there. 
Consequently the Bose condensation temperature for the former system
would be higher than that for the latter.
The growth of the condensate fraction for different
lattices is shown in Figs. (1)-(3), and in Fig (4) we have
compared condensate fraction growth for different lattices.
Also shown by dotted lines is the condensate fraction for
free bosons and they are found to be rather close.
The variation of $T_B$ with $n$ is shown in Fig. (5) which
shows an initial fast growth and a monotonic increase for
higher values of $n$.
\subsection{Weakly interacting bosons}
\label{subsec}
To study Bose condensation of weakly interacting Bosons in tight
binding bands, we employ the following Hamiltonian:
\begin{eqnarray}
\displaystyle
H&=&\sum_{{\bf k}}\left[\epsilon({\bf k})-\mu \right]
c^{\dag}_{{\bf k}}c_{{\bf k}}
\nonumber \\
&+& \frac{U}{2N_{s}}
\sum_{{\bf k}}\sum_{{\bf k}^{\prime}}\sum_{{\bf q}}
c^{\dag}_{{\bf k+q}}c^{\dag}_{{\bf k^{\prime}-q}}
c_{{\bf k}^{\prime}}c_{{\bf k}}\,.
\end{eqnarray}
Here $\epsilon({\bf k})$ is the boson band structure, $\mu$ the
chemical potential, $c^{\dag}_{\bf k}$ the boson creation 
operator, 
$U$ the boson-boson repulsive interaction energy
taken to be a constant for simplicity,
and $N_s$ the number of lattice sites. In this section, we will
deal with a range of $U$ such that $U\,<\,2W$ where $W$ is the
half-band-width. Our aim is to get a boson number equation in terms of
$\epsilon({\bf k})$, $U$, $\mu$, and temperature. One can then calculate
the condensate fraction and the transition temperature.
To this end we follow Bogoliubov theory\cite{bogo}. In this theory,
one makes an {\em assumption} that the ground state of interacting bosons
is a Bose condensate.
Since the lowest single particle state (which is 
${\bf k}=0$ in our case of simple tight binding bands) has macroscopic
occupation (say $N_0$), we have $<c^{\dag}_{0}c_{0}>\, 
\approx\, <c_{0}c^{\dag}_{0}>$.
Then the the operators $c^{\dag}_{0}$ and $c_{0}$ can be treated\
as complex numbers and  one gets $<c^{\dag}_{0}>$ = $<c_{0}>$ = $\sqrt{N_0}$.
This complex number substitution has been recently shown\cite{lieb}
to be justified.
Clearly we have a two-fluid system consisting of two subsystems of 
condensed and non-condensed
bosons. There are interactions between particles within each subsystem
and interaction between particles in the two subsystems. In the Bogoliubov
approach, to obtain second order interaction terms one makes the 
substitution: $c^{\dag}_{0}\rightarrow \sqrt{N_0}+c^{\dag}_{0}$. 
In the ground state, the linear fluctuation terms terms must vanish
and this fixes the chemical potential to $\mu\,=\,Un_{0}+\epsilon_{0}$,
where $\epsilon_{0}$ is the minimum of the single particle
energy spectrum (which is equal to $-zt$ for a bi-partite lattice
with co-ordination number $z$), and $n_{0}\,=\,N_{0}/N_{s}$. Following
Bogoliubov approach,
after a mean-field factorization of the interaction term, we obtain
for lattices with inversion symmetry the following  mean-field
Hamiltonian:
\begin{eqnarray}
\displaystyle
H_{BMF}&=&-E_{0}+\sum^{\prime}_{{\bf k}}\frac{[\xi({\bf k})+Un_{0}]}{2}
(c^{\dag}_{{\bf k}}c_{{\bf k}}+c^{\dag}_{{\bf -k}}c_{{\bf -k}})
\nonumber \\
&+&\frac{Un_{0}}{2}\sum^{\prime}_{{\bf k}}
(c^{\dag}_{{\bf k}}c^{\dag}_{{\bf -k}}+c_{{\bf -k}}c_{{\bf k}})\,,
\end{eqnarray}
where $E_{0}\equiv -Un_{0}N_{0}/2$ and 
$\xi({\bf k})\equiv \epsilon({\bf k})-\epsilon_{0}$.
As mentioned earlier our aim is to get an equation for number
of particles. This can be obtained from the Green's function\cite{zubarev}
$G({\bf k},\omega)\equiv 
<<c_{{\bf k}};c^{\dag}_{{\bf k}}>>_{\omega}$ using the 
relation:
\begin{equation}
n_{{\bf k}}=\lim_{\eta \rightarrow 0}\int_{-\infty}^{\infty}
\left[G({\bf k},\omega+i\eta)-G({\bf k},\omega-i\eta)\right]\,
f(\omega)d\omega\,.
\end{equation}
where $f(\omega)=1/[exp(\omega/k_{B}T)-1]$.
The Heisenberg equation of motion for $G({\bf k},\omega)$ 
is:
\begin{equation}
\omega\,G({\bf k},\omega)=[c_{{\bf k}},c^{\dag}_{{\bf k}}]
+<<[c_{{\bf k}},H];c^{\dag}_{{\bf k}}>>_{\omega}\,,
\end{equation}
where $H$ is the Hamiltonian of the system. Using $H_{BMF}$,
we obtain:
\begin{equation}
\omega\,G({\bf k},\omega)=
1+[\xi({\bf k})+Un_{0}]G({\bf k},\omega)+Un_{0}F({\bf k},\omega)\,,
\end{equation}
and the Green's function to which $G({\bf k},\omega)$ is
coupled $F({\bf k},\omega)\equiv 
<<c^{\dag}_{{\bf -k}};c^{\dag}_{{\bf k}}>>_{\omega}$
obeys the equation of motion:
\begin{equation}
\omega F({\bf k}, \omega)=-[\xi({\bf k})+Un_{0}]F({\bf k}, \omega)
-Un_{0}G({\bf k}, \omega).
\end{equation}
Solving the above two equations one obtains:
\begin{eqnarray}
\displaystyle
G({\bf k}, \omega)&=&
\left(
-\frac
{\omega+\xi({\bf k})+Un_{0}} {2E({\bf k})}
\right) \nonumber \\
&\times &
\left(
\frac{1}{\omega+E({\bf k})}-\frac{1}{\omega-E({\bf k})}
\right)\,.
\end{eqnarray}
where the Bogoliubov quasiparticle energy $E_{{\bf k}}$ is:
\begin{equation}
E({\bf k})=\sqrt{\xi^{2}({\bf k})+2Un_{0}\xi({\bf k})}.
\end{equation}
Note that for the tight binding band dispersions used, $E({\bf k})$
is linear in ${\bf k}$ in the long wave-length limit.
Now, the number of particles per site ($n$) is readily obtained
using Eqs. (7) and (11) to be:
\begin{eqnarray}
\displaystyle
n&=&n_{0}+\frac{1}{2N_{s}}\sum^{\prime}_{{\bf k}}
\left[
\left(1+\frac{\xi({\bf k})+Un_{0}}{E_{{\bf k}}}\right)
\frac{1}{e^{\beta E({\bf k})/k_{B}T}-1}
\right]
\nonumber \\
&+&\frac{1}{2N_{s}}\sum^{\prime}_{{\bf k}}
\left[
\left(1-\frac{\xi({\bf k})+Un_{0}}{E_{{\bf k}}}\right)
\frac{1}{e^{-\beta E({\bf k})/k_{B}T}-1}
\right].
\end{eqnarray}
It is useful to look at some limits of the above equation.
When $U=0$ and $T=0$, we have $n_{0}=n$ which means that all the
particles are in the condensate at absolute zero in the non-
interacting limit. When $U=0$ and $T\neq 0$, on obtains:
\begin{equation}
n_{0}=n-\frac{1}{N_{s}}\sum^{\prime}_{{\bf k}}
\frac{1}{e^{\frac{\xi({\bf k})}{k_{B}T}}-1}\,.
\end{equation}
The second term on the RHS of the above
equation is the thermal depletion of the condensate.
Further, when $U \neq 0$ and $T=0$, we get:
\begin{equation}
n_{0}=n-\frac{1}{N_{s}}\sum^{\prime}_{{\bf k}}
\left(
\frac{\xi({\bf k})+Un_{0}-E({\bf k})}{2E({\bf k})}
\right)\,,
\end{equation}
in which second term on the RHS is
the interaction induced depletion of the condensate.
The interaction has twin effects of leading to
a modified excitation
spectrum and to a partial depletion of the condensate.
The gap-less and linear long-wavelength excitation spectrum is consistent with
experimental measurements\cite{greiner} on interacting bosons
in their bose-condensed state in optical
lattices. 
In Fig. (6), we have shown the numerical solution of 
Eq. (13) for  {\em bcc, fcc} and {\em sc} lattices.
The condensate fraction ($n_{0}/n$) is
seen to be gradually suppressed with increasing $U/W$.
Note also that at $U=0$, all the particles are in the condensate. 
Finally, we consider the case: $U\neq 0$ and $T \neq 0$. 
In Fig. (7), we have 
displayed the variation of the condensate fraction as a function
of temperature for various values of $U/W$.
The interaction partially depletes the condensate 
at zero temperature and close to it,
while enhancing it beyond this range below the
Bose-Einstein condensation temperature.
We have not plotted these curves
all the way to $T_{B}$ since the Bogoliubov approximation breaks down
close to $T_{B}$. The variation of $T_B$ with $n$ is same as in the
case of non-interacting bosons as can be seen by setting $n_{0}=0$
in Eq. (13).
At very low temperature, thermal depletion is negligible and correlation
induced depletion causes a reduction of $n_0$ with increasing $U$. At higher
temperatures when thermal depletion is important, $U$ plays another role.
The energies of the excited states shift to larger values with
increasing $U$, consequently the population in the excited states decreases
and an enhancement of $n_0$ occurs with increasing $U$.
\par
The analysis presented in this section is reasonable provided
the effect of interaction  is perturbative. 
When the interaction strength increases, there is a possibility
for a correlation induced localization transition for  interacting 
bosons in a narrow band.
In the next section, we analyze this strongly
interacting bosons case. 
\subsection{Strongly interacting bosons}
\label{subsec}
We first write the Hamiltonian, Eq. (5), in real space. Then we have:
\begin{equation}
H= \sum_{ij}[-t-\mu \delta_{ij}]c^{\dag}_{i}c_{j}
+\frac{U}{2}\sum_{i}n_{i}(n_{i}-1)\,,
\end{equation}
where $n_{i}=c^{\dag}_{i}c_{i}$. 
For simplicity, let us confine to the case of
$n \leq 1$. The effect of increasing interaction ($U$) is to make
motion of the bosons in the lattice correlated so as to 
avoid multiple occupancy of the sites. In the dilute limit, 
the effect of $U$ is not serious since there are enough
vacant sites. The effect of $U$ then is prominent when
$n$ is close to unity or to an integer value in the general case.
In the large $U$ limit, it becomes
favorable for the bosons to localize on the sites to avoid
the energy penalty of multiple site occupancy. Qualitatively then,
one can see that increasing $U$ increases the effective mass
(or decreases the band-width)
of the bosons and eventually drives them, for integer filling,
to a BMH insulator state. 
For large $U$, when the double or multiple occupancy is forbidden
one can calculate the band-width reduction factor
[$\phi_{B}(n)$] approximately following 
the spirit of Renormalized Hamiltonian 
Approach (RHA) to fermion Hubbard model\cite{zhang,das} based on 
Gutzwiller approximation\cite{gutz,voll}.
Within the RHA, the effect of projecting out double or
multiple occupancies on a non-interacting boson wave function is taken
into account by a 
a classical renormalization factor which is the ratio
of the probabilities of the corresponding physical process
in the projected and unprojected spaces.
The probability of a hopping process in the projected space
is given by $n(1-n)$ for $n \leq 1$. This simply implies
that the site from which hopping takes place must be occupied
and the target site must be empty in the projected space.
In the unprojected space, the probability of hopping
is just equal to the probability of the site from which
hopping takes place being occupied. The hopping takes
place for non-interacting bosons irrespective of the target site being
empty or occupied by any number of bosons.
This probability may be found out by calculating
the number of ways a given number of non-interacting bosons is 
distributed in $N_S$ number of lattice sites and $(N_{S}-1)$
number of lattice sites. The difference would give the number
of configurations where a particular site is occupied.
Following this route, the probability that a site is 
occupied is obtained as:
\begin{equation}
p(N;N_{s})=1- \frac{(N+N_{s}-2)!(N_{s}-1)!}{(N_{s}-2)!(N+N_{s}-1)!},
\end{equation}
where $N$ and $N_s$ are the total number of bosons and lattice sites,
respectively. In the thermodynamic limit:
\begin{equation}
p(n)=\frac{n}{1+n}\,.
\end{equation}
So, the hopping probability is just $p(n)$.
The above equation is valid for any $n$. Now, in the strongly correlated state
(large $U$ limit), confining to the case of $n\leq 1$, the hopping probability
is $n\,(1-n)$. Hence the $\phi_{B}(n)$ is obtained to
be, 
\begin{equation}
\phi_{B}(n)=1-n^{2}.
\end{equation}
The above equation is valid only for $n \leq 1$ and in the large $U$
limit. It may be interesting for the reader to note that, in
the Fermion case, 
$\phi_{F}(n)=2(1-n)/(2-n)$ which has been used in the studies\cite{and} of
superconductivity in strong coupling fermion Hubbard model
of high temperature superconductors. Since in the
large $U$ limit, double or higher site-occupancies are forbidden,
one can write a renormalized Hamiltonian, valid for
$n \leq 1$, for strongly correlated
bosons as :
\begin{equation}
H_{sc}=\sum_{{\bf k}}[\phi_{B}(n)\epsilon({\bf k})-\mu]
c^{\dag}_{{\bf k}}c_{{\bf k}}.
\end{equation} 
The above $H_{sc}$ is clearly the Hamiltonian of {\em non-interacting}
bosons in a narrow band which has undergone a strong
correlation induced filling-dependent band-narrowing.
One can see that as $n$ increases from $0$ to $1$, the boson
effective mass increases to eventually diverge at $n=1$ and
a BMH insulator obtains. We do admit that there are limitations
to this renormalized Hamiltonian approach. While 
it has a merit that
detailed
band structure information can be incorporated in 
the Bose condensation temperature 
calculation, it has the demerit that 
we have to restrict ourselves to  large $U$ and $n \leq 1$. 
The above Hamiltonian  is valid for $U/W\,>\,(U/W)_{c}$ where $(U/W)_{c}$
is the critical
value  for transition into the Mott insulating phase for n = 1.
Fixing a precise lower limit on $(U/W)$ is not possible in the absence
of either exact analytical or numerical solution of three dimensional
Bose-Hubbard model.
It may be mentioned that in this large $U$ limit and for $n \leq 1$, the 
Bose-Hubbard model is reduced to the lattice Tonks (hard-core boson) 
gas
which is different from the classical gas of elastic hard
spheres investigated by Tonks\cite{tonks}. 
Our results imply then that the effective mass of 
bosons in a lattice Tonks gas in a narrow energy band 
is strongly band-filling 
dependent.
The variation of bose condensation temperatures  with $n$ for bosons with 
correlation-induced renormalized energy bands 
corresponding to {\em sc, bcc}, and {\em fcc} lattices
are displayed in Fig. (8). In the region between each curve and
the $n$-axis, the bosons are in their bose-condensed state, and above
each curve they are in their normal state (except at
$n=1$ and below $k_{B}T\approx U$).
The variation of $T_{B}$ is a
resultant of the combined effects of increasing density and
increasing effective mass since $T_{B}$ is proportional
to $n/m^{*}$. Beyond around twenty percent filling, the 
increasing effective mass over-compensates the effect of
increasing $n$ and pulls down the growth of $T_{B}$ eventually
driving it to zero at $n=1$ at which density one has a BMH
insulator.
\par 
It is of some interest to make a comparison between Mott-Hubbard
(MH) metal-insulator transitions observed in fermion systems in 
condensed matter physics. For a half-filled narrow band of fermions,
one way to induce a MH insulator to metal transition is by
reducing the ratio of Coulomb repulsion ($U$) to the band-width
($2W$). This was experimentally achieved\cite{rice} in $V_{2}O_{3}$ by
application of pressure. This then is a band-width controlled
MH transition\cite{brink}.
Another way to induce a MH insulator to metal transition
is to start with a Mott insulator and reduce the band-filling which
would then be filling-controlled MH transition. This was achieved
\cite{tokura} in
$La_{x}Sr_{1-x}TiO_{3}$. 
For a simple Gutzwiller approximation based analysis of the properties of
this material see Ref. \onlinecite{ram}.
Now, the Mott transition observed\cite{greiner} in
boson systems in optical lattices is the band-width controlled one.
It would be interesting to look for filling-controlled Mott transition
in boson systems in optical lattices which may be possible by 
starting with the BMH insulator and flipping
a few atoms out of the trap.
\section{Conclusions}
\label{sec3}
In this paper we presented a theoretical study of
condensation of bosons in tight binding bands corresponding
to {\em sc}, {\em bcc}, and
{\em fcc} lattices. We analyzed condensation
temperature and condensate
fraction of 
non-interacting bosons, weakly interacting bosons using Bogoliubov
method, and strongly interacting bosons through a renormalized
Hamiltonian approach (limited to $n \leq 1$) 
capable of incorporating the detailed
boson band structures.
In all the cases studied, we find that
bosons in a tight binding band corresponding to a 
{\em bcc} lattice has the highest Bose condensation
temperature. The growth of condensate fraction of non-interacting
bosons is found to be very close to that of free bosons. 
In the case of weakly interacting bosons, the
interaction partially depletes the condensate 
at zero temperature and close to it,
while enhancing it beyond this range below the
Bose-Einstein condensation temperature.
Strong interaction enhances
the boson effective mass as the band-filling is increased and
eventually localizes them to form a Bose-Mott-Hubbard insulator
for $n=1$. In the strongly interacting bosons case, we found that
all bosons are in the condensate at absolute zero temperature.
We also pointed out a possibility of
a filling-controlled BMH transition for bosons in optical 
lattices.
\acknowledgments
We thank Prof. R. K. Moitra for many useful discussions. 
We also thank Prof. Bikas Chakrabarti for a useful discussion.

\begin{figure}
\resizebox*{3.1in}{2.5in}{\rotatebox{270}{\includegraphics{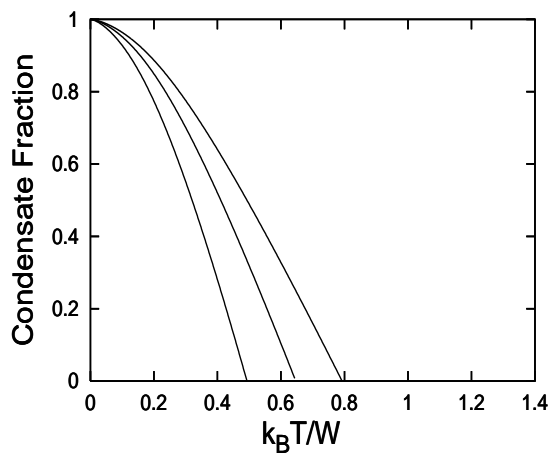}}}
\vspace*{0.5cm}
\caption[]{
The varation of condensate fraction with temperature for bosons
in {\em sc} lattice for $n\,=\,0.8$ (top),
$n\,=\,0.6$ (middle), $n\,=\,0.4$ (bottom).
In this and other figures $W$ is the half-band-width.}
\label{scaling}
\end{figure}
\begin{figure}
\resizebox*{3.1in}{2.5in}{\rotatebox{270}{\includegraphics{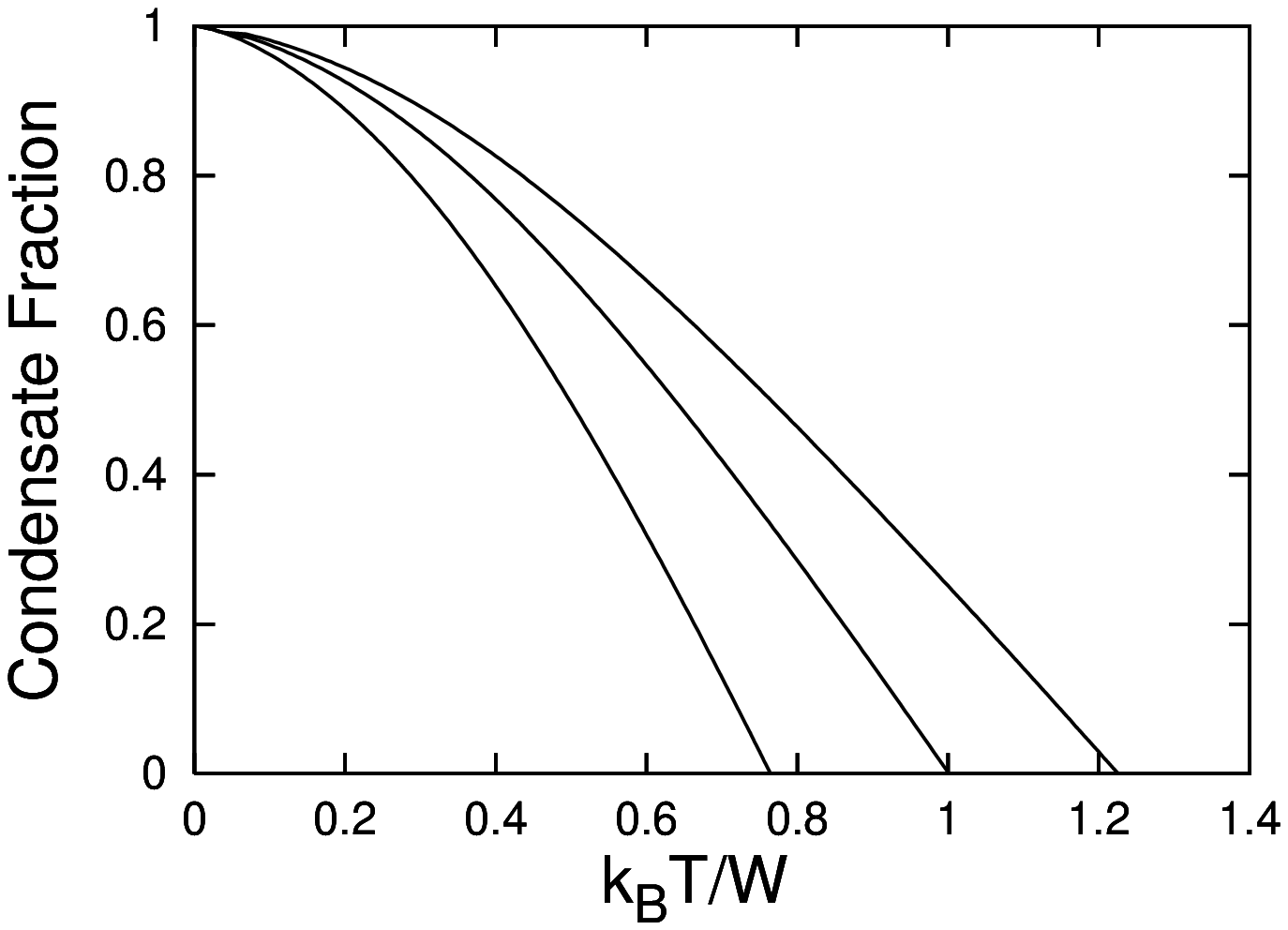}}}
\vspace*{0.5cm}
\caption[]{
The same as in Fig. 1 for bosons
in a {\em bcc} lattice for $n\,=\,0.8$ (top),
$n\,=\,0.6$ (middle), $n\,=\,0.4$ (bottom).
}
\label{scaling}
\end{figure}
\begin{figure}
\resizebox*{3.1in}{2.5in}{\rotatebox{270}{\includegraphics{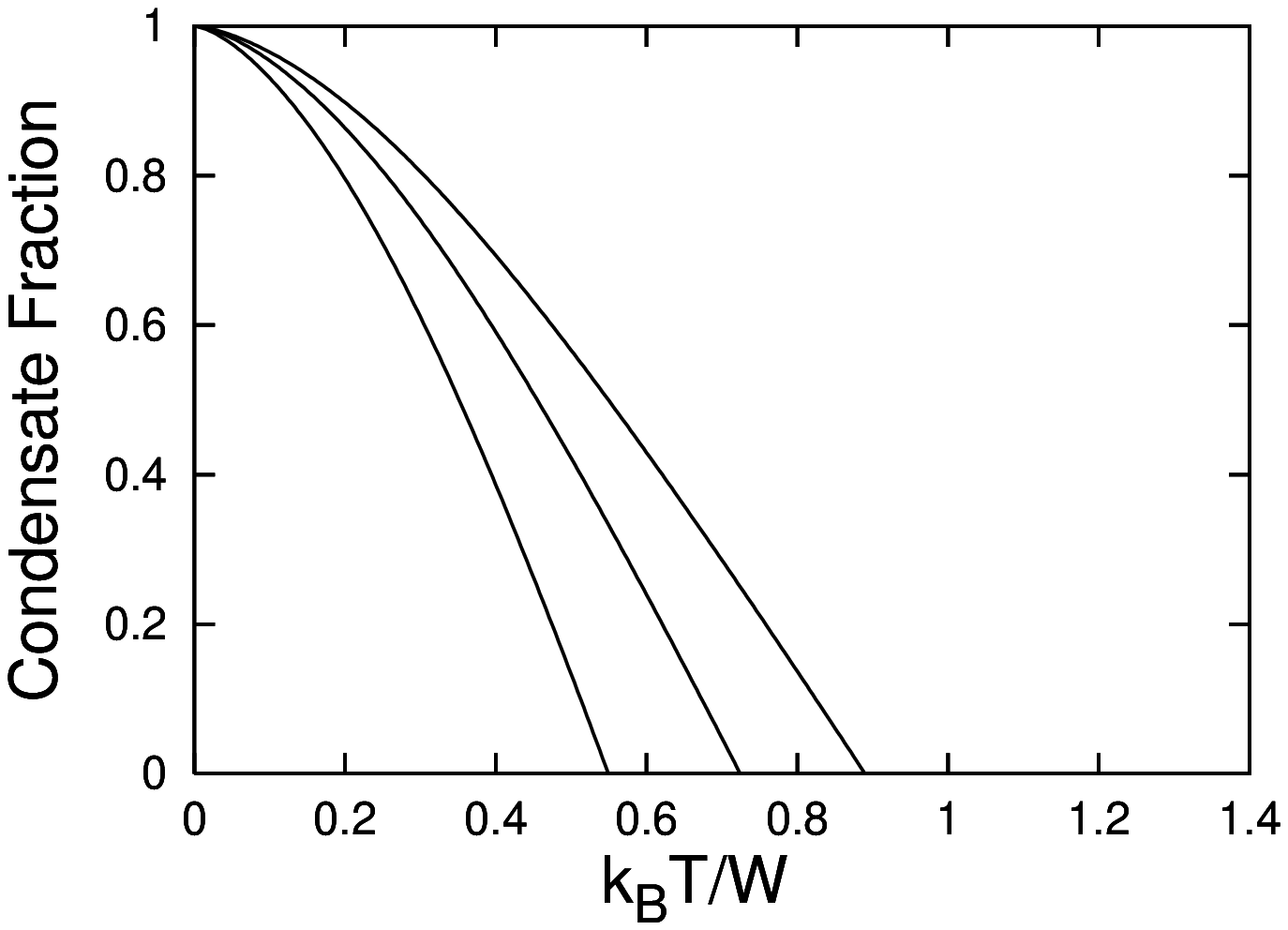}}}
\vspace*{0.5cm}
\caption[]{
The same as in Fig. 1 for bosons
in a {\em fcc} lattice for $n\,=\,0.8$ (top),
$n\,=\,0.6$ (middle), $n\,=\,0.4$ (bottom).
}
\label{scaling}
\end{figure}
\begin{figure}
\resizebox*{3.1in}{2.5in}{\rotatebox{270}{\includegraphics{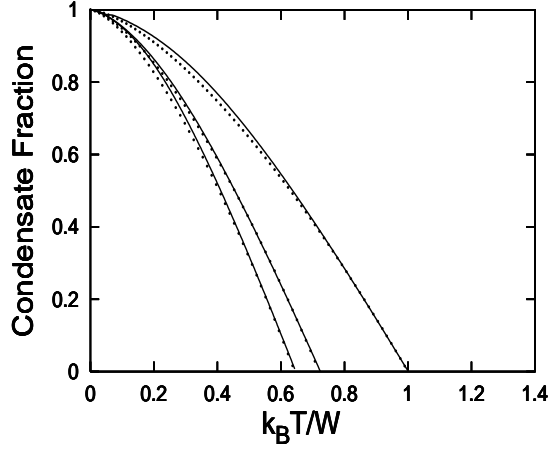}}}
\vspace*{0.5cm}
\caption[]{
The variation of condensate fraction (for $n\,=\,0.6$) for 
{\em bcc} (top), {\em fcc} (middle), and {\em sc} (bottom) lattices.
The dots are plots of $1-(T/T_{c})^{3/2}$.
}
\label{scaling}
\end{figure}
\begin{figure}
\resizebox*{3.1in}{2.5in}{\rotatebox{270}{\includegraphics{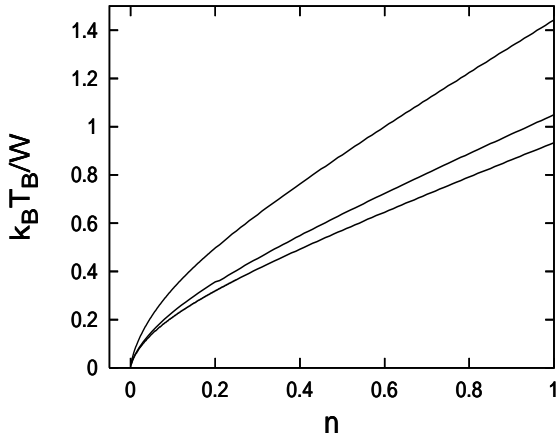}}}
\vspace*{0.5cm}
\caption[]{
The variation of Bose condensation temperature 
(of non-interacting or weakly interacting bosons) with $n$ for
{\em bcc} (top), {\em fcc} (middle), and {\em sc} (bottom) lattices.
}
\label{scaling}
\end{figure}
\begin{figure}
\resizebox*{3.1in}{2.5in}{\rotatebox{270}{\includegraphics{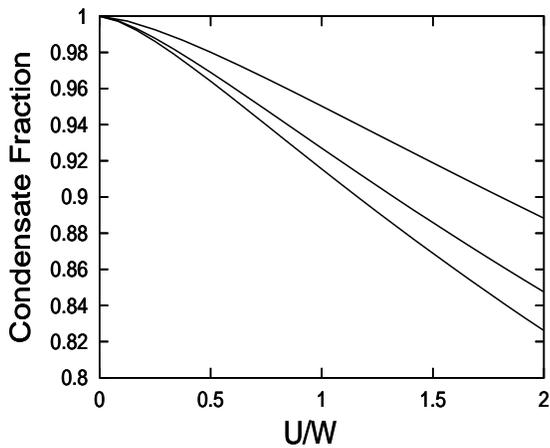}}}
\vspace*{0.5cm}
\caption[]{
The variation of condensate fraction (for $T\,=\,0$ and 
$n\,=\,0.4$) with $U/W$ for 
{\em bcc} (top), {\em fcc} (middle), and {\em sc} (bottom) lattices.
}
\label{scaling}
\end{figure}
\begin{figure}
\resizebox*{3.1in}{2.5in}{\rotatebox{270}{\includegraphics{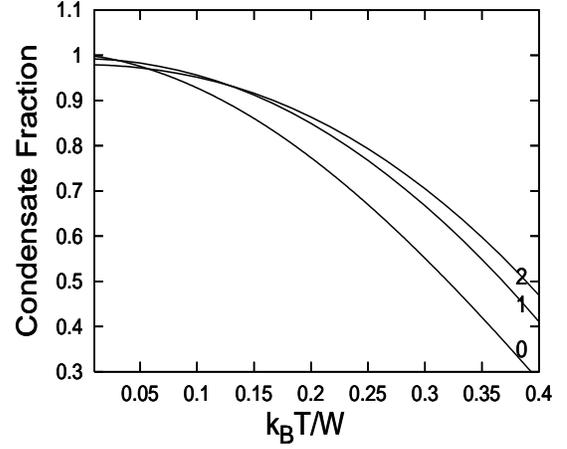}}}
\vspace*{0.5cm}
\caption[]{Condensate fraction (for $n\,=\,0.4$) vs 
temperature for various values
of $U/W$ (shown on the curves) and for {\em sc} lattice.
}
\label{scaling}
\end{figure}
\begin{figure}
\resizebox*{3.1in}{2.5in}{\rotatebox{270}{\includegraphics{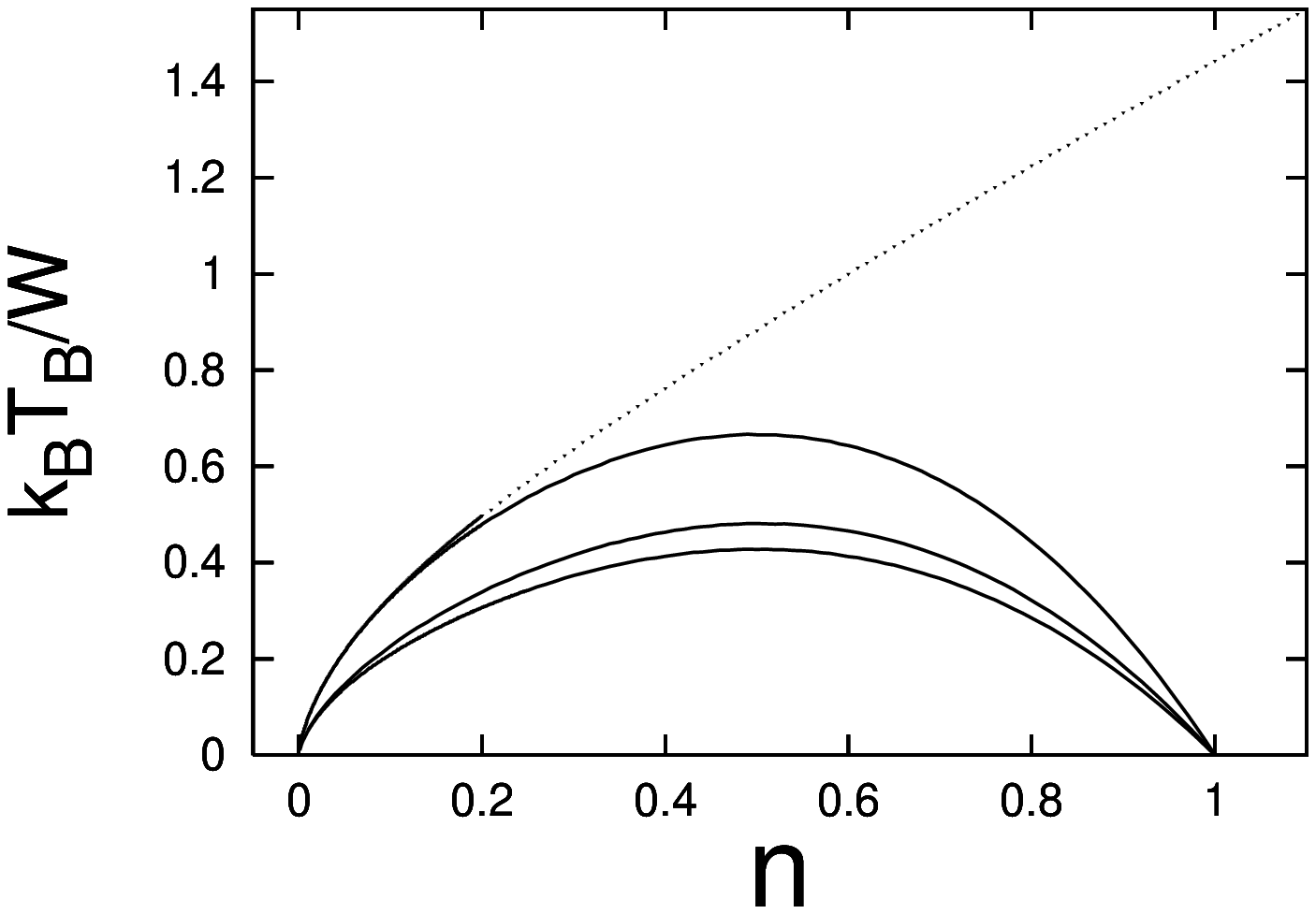}}}
\vspace*{0.5cm}
\caption[]{Bose condensation temperature 
(of strongly
interacting bosons) vs $n$ for
{\em bcc} (top), {\em fcc} (middle), and {\em sc} (bottom) lattices.
The dots are for non-interacting bosons in a {\em bcc} lattice.}
\label{scaling}
\end{figure}
\end{multicols}

\begin{references}
\bibitem[b] {add2} on {\em lien} from SINP, Kolkata.
\bibitem{einstein}
A. Einstein, Sitzber. Kgl. Preuss. Akad. Wiss. 261 (1924); 3(1925).
\bibitem{bose}
S. N. Bose, Z. Phys. {\bf 26}, 178 (1924).
\bibitem{cornell}
M. H. Anderson, J. R. Ensher, M. R. Matthews, C. E. Weiman, and
E. A. Cornell, Science {\bf 269}, 198 (1995).
\bibitem{davis}
K. B. Davis, M. -O. Mewes, M. R. Andrews, N. J. van Druten, 
D. S. Durfee, D. M. Kurn, and W. Ketterle, 
Phys. Rev. Lett. {\bf 75}, 3969 (1995).
\bibitem{bradley}
C. C. Bradley, C. A. Sackett, J. J. Tollett, and R. G. Hulet, Phys. Rev.
Lett. {\bf 75}, 1687 (1995).
\bibitem{klep}
D. G. Fried, T. C. Killian, L. Willmann, D. Landhuis, S. C. Moss,
D. Kleppner, and T. J. Greytak, Phys. Rev.
Lett. {\bf 81}, 3811 (1998).
\bibitem{pita}
F. Dalfovo, S. Giorgini, L. P. Pitaevskii, S. Stringari,
Rev. Mod. Phys. {\bf 71}, 463 (1999).
\bibitem{leg}
A. J. Leggett, Rev. Mod. Phys. {\bf 73}, 307 (2001).
\bibitem{ketterle}
J. R. Anglin and W. Ketterle, Nature {\bf 416}, 211 (2002).
\bibitem{pethick}
C. J. Pethick and H. Smith, {\em Bose-Einstein 
condensation in dilute gases} (Cambridge University Press, Cambridge UK, 2002). 
\bibitem{hall}
D. S. Hall. Am. J. Phys. {\bf 71}, 649 (2003).
\bibitem{greiner}
M. Greiner, O. Mandel, T. Esslinger, T. W. H\"{a}nsch, and I. Bloch,
Nature {\bf 415}, 39 (2002).
\bibitem{schori}
C. Schori, T. St\"{o}ferle, H. Moritz, M. K\"{o}hl, T. Esslinger,
Phys. Rev. Lett. {\bf 93}, 240402 (2004).
\bibitem{jak}
D. Jaksch, C. Bruder, J. I. Cirac, C. W. Gardiner, and P. Zoller,
Phys. Rev. Lett. {\bf 81}, 3108 (1998).
\bibitem{fisher}
M. P. A. Fisher, P. B. Weichman, G. Grinstein, and D. S. Fisher, Phys. Rev.
B {\bf 40}, 546 (1989).
\bibitem{nandini}
W. Krauth and N. Trivedi, Europhys. Lett. {\bf 14}, 627 (1991).
\bibitem{shesh}
K. Sheshadri, H. R. Krishnamurthy, R. Pandit, and T. V. Ramakrishnan,
Europhys. Lett. {\bf 22}, 257 (1993).
\bibitem{monien}
J. K. Freericks and H. Monien, Phys. Rev. B {\bf 53}, 2691 (1996);
N. Niemeyer, J. K. Freericks, and H. Monien, Phys. Rev. B {\bf 60}, 2357 (1999).
For recent work in this direction see
P. Buonsante, V. Penna, and A. Vezzani, Phys. Rev. B {\bf 70}, 184520 (2004).
\bibitem{oost}
D. Van Oosten, P. van der Straten, and H. T. C. Stoof, Phys. Rev.
A {\bf 63}, 054601 (2001).
\bibitem{stof}
T. St\"{o}ferle, H. Moritz, C. Schori, M. K\"{o}hl, T. Esslinger,
Phys. Rev. Lett. {\bf 92}, 130403 (2004).
\bibitem{black}
P. B. Blakie and C. W. Clark, J. Phys. {\bf B} 37, 1391 (2004).
\bibitem{baizakov}
B. B. Baizakov, M. Salerno, and B. A. Malomed, in {\em Nonlinear Waves:
Classical and Quantum Aspects}, edited by F. Kh. Abdullaev and
V. V. Konotop (Kluwer Academic Publishers, Dordrecht, 2004).
\bibitem{santos}
L. Santos, M. A. Baranov, J. I. Cirac, H. U. Everts, H. Fehrmann, and M Lewenstein,
Phys. Rev. Lett. {\bf 93}, 030601 (2004).
\bibitem{bogo}
N. N. Bogoliubov, J. Phys. (Moscow) {\bf 11}, 23 (1947).
\bibitem{lieb}
E. H. Lieb, R. Seiringer, and J. Yngvason, Phys. Rev. Lett. {\bf 94},
080401(2004).
\bibitem{zubarev}
D. N. Zubarev, Usp. Fiz. Nauk. {\bf 71}, 71 (1960) 
[Sov. Phys.-Usp. 3, 320 (1960)]. 
\bibitem{zhang}
F. C. Zhang, C. Gros, T. M. Rice, and H. Shiba,
Supercond. Sci. Technol. {\bf 1}, 36(1988).
\bibitem{das}
A. N. Das, J. Konior, A. M. Oles, and D. K. Ray, Phys. Rev. B {\bf 44}, 
7680 (1991).
\bibitem{gutz}
M. C. Gutzwiller, Phys. Rev. Lett. {\bf 10}, 159 (1963).
\bibitem{voll}
D. Vollhardt, Rev. Mod. Phys. {\bf 56}, 99 (1984).
\bibitem{and}
P. W. Anderson, P. A. Lee, M. Randeria, T. M. Rice,
N. Trivedi, and F. C. Zhang, J. Phys. CM {\bf 16}, R755 (2004).
\bibitem{tonks}
L. Tonks, Phys. Rev. {\bf 50}, 955 (1936).
\bibitem{rice}
D. B. McWhan, A. Menth, J. P. Remeika, W. F. Brinkman, and T. M. Rice,
Phys. Rev. B {\bf 7}, 1920 (1973).
\bibitem{brink}
W. F. Brinkman and T. M. Rice, Phys. Rev. B {\bf 2}, 4302 (1970).
\bibitem{tokura}
Y. Tokura, Y. Taguchi, Y. Okada, Y. Fujishima, T. Arima, K. Kumagai,
and Y. Iye, Phys. Rev. Lett. {\bf 70}, 2126 (1993).
\bibitem{ram}
R. Ramakumar, K. P. Jain, and C. C. Chancey, 
Phys. Rev. B {\bf 50}, 10122 (1994).
\end{references}
\end{document}